\documentclass[11pt]{article}

\usepackage{coling2020}
\usepackage{times}
\usepackage{url}
\usepackage{soul}

\usepackage{graphicx}
\usepackage{booktabs}
\usepackage{algorithm}
\usepackage{algorithmic}
\usepackage{CJKutf8}
\usepackage{subfigure}
\usepackage{xcolor}
\usepackage{amsmath,amsthm,bm}
\usepackage{multirow}
\def \todo #1 {\textcolor{red}{TODO: #1}}
\def \ls #1 {\textcolor{olive}{LS: #1}}
\def \grj #1 {\textcolor{blue}{GRJ: #1}}

\title{APPCorp: A Corpus for Android Privacy Policy Document Structure Analysis}

\author{
  Shuang Liu \and Renjie Guo \and Baiyang Zhao\\
  School of Intelligence and Computing \\
  Tianjin University \\
  {\tt shuang.liu@tju.edu.cn} \\
  {\tt guorenjie@tju.edu.cn} \\
  {\tt zhaobaiyang@tju.edu.cn}
  \AND
  Tao Chen \\
  Johns Hopkins University \\
  {\tt chentaokite@gmail.com} \\ \And
  Meishan Zhang \\
  School of New Media and Communication\\
  Tianjin University \\
  {\tt mason.zms@gmail.com} \\ \AND
  }

\date{}

\begin{document}
\maketitle
\begin{abstract}
With the increasing popularity of mobile devices and the wide adoption of mobile Apps, an increasing concern of privacy issues is raised.
Privacy policy is identified as a proper medium to indicate the legal terms, such as GDPR, and to bind legal agreement between service  providers and users. However, privacy policies are usually long and vague for end users to read and understand. It is thus important to be able to automatically analyze the document structures of privacy policies to assist user understanding.
In this work we create a manually labelled corpus containing $167$ privacy policies (of more than $447$K words and $5,276$ annotated paragraphs).
We report the annotation process and details of the annotated corpus.
We also benchmark our data corpus with $4$ document classification models, thoroughly analyze the results and discuss challenges and opportunities for the research committee to use the corpus. We release our labelled corpus as well as the classification models for public access.
\end{abstract}


\section{Introduction}
\label{sec:introduction}
\begin{figure}[tp]
\centering
\begin{subfigure}{12.7cm}
\includegraphics[width=\linewidth]{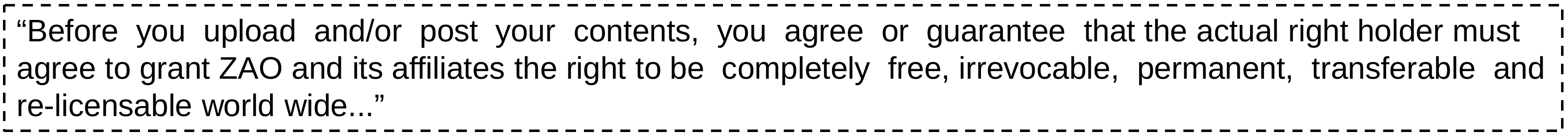}
\caption{The ZAO Privacy Policy Excerpt in English translation}\label{fig:zao}
\label{fig:motivationa}
\end{subfigure}\qquad

\begin{subfigure}{12.7cm}
\includegraphics[width=\linewidth]{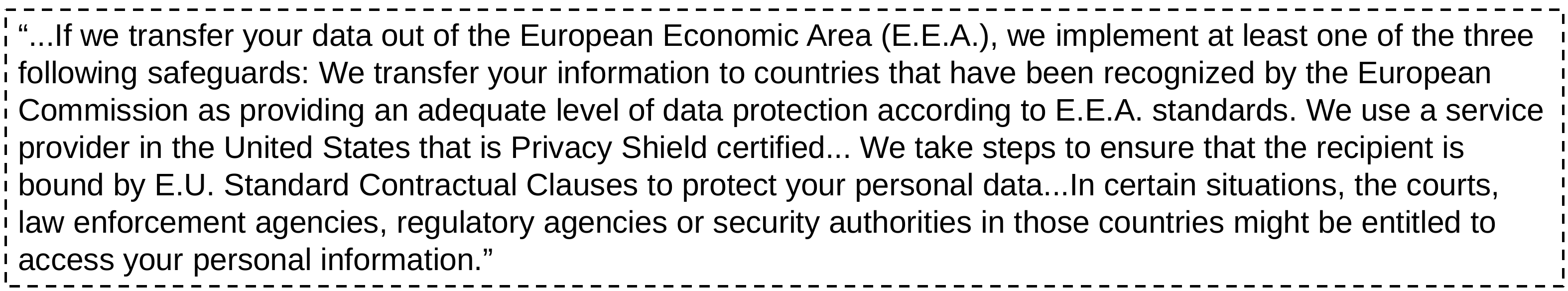}
\caption{The New York Times Privacy Policy Excerpt}\label{fig:nyt}
\label{fig:motivationb}
\end{subfigure}
\label{fig:motivation}
\caption{The Privacy Policy Excerpt Examples}
\end{figure}

With the rapid development of mobile applications and their wide adoption in different domains, more and more personal data has been provided to application providers.
Privacy policy is a document which binds the legal agreement between service providers and users.
Therefore, it is fairly important for users to understand the contents of privacy policies before they tick the ``I agree'' box.
For instance, the privacy policy of the ZAO, a Chinese deepfake-like application, explicitly states that the ownership of users' personal data (in particular images uploaded to ZAO) are unconditionally and permanently transferred to ZAO and its affiliates\footnote{The information is extracted from news snapshot, the privacy policy of ZAO has been updated after the report.}~(excerpt shows in Figure~\ref{fig:zao}). However, many users of the service were unaware of this term on agreeing the privacy policy.

Reading privacy policies is extremely time consuming: each American Internet user needs to spend $244$ hours per year to read all the online privacy policies of her visited sites~\cite{ppreadingcost}.
This is a common, yet hard-to-solve issue due to three reasons.
1) Privacy policies are usually long documents, which are time consuming to read. For instance, on average every privacy policy document has $2,677$ words in our dataset.  2) Some privacy policies are poorly structured, or with no structure at all, which makes it harder to read. 3) Privacy policies are usually written 
with legal terms that are hard for non-experts to understand.


Several studies have been conducted to perform automatic/semi-automatic analysis on privacy policies~\cite{liu2018towards} and human labeled corpora~\cite{PPCorpus16,zimmeck2019maps} are created for the purpose. However, existing work focus on specific and fine-grained aspects of privacy policies, such as vague words~\cite{VagueCorpus18}, detailed information types~\cite{zimmeck2019maps}, fine-grained attributes for text segments~\cite{PPCorpus16}. None of them attempt to uncover the topics of paragraphs, which could be very useful for outlining or restructuring long documents.

Moreover, regulations have huge impacts on privacy policies~\cite{kaur2018comprehensive}, especially for the enacting of General Data Protection Regulation (GDPR)\footnote{\url{https://gdpr-info.eu/}}. For instance, The \texttt{New York Times} privacy policy was updated on May 24th, 2018 (one day before GDPR was formally put into action) and a term specifying \texttt{International Data Transfer}, which is explicitly required in GDPR, was added. Figure~\ref{fig:nyt} shows the corresponding excerpt of the updated privacy policy. Similarly, $53$ out of the $115$\footnote{There are $10$ privacy policies which are not accessible.} privacy policies in the OPP-$115$ corpus~\cite{PPCorpus16} are changed due to GDPR related regulations. A large-scale study~(\newcite{sarne2019unsupervised}) found mismatches of the topics extracted from privacy policies (after GDPR is enacted) with the OPP-115 labels. However, none of the existing work considers the legislation impacts on privacy policies.

In this work, we propose a novel task of categorizing the topical structures of privacy policies for Android Apps.  We devise a classification scheme to characterize the topics for paragraphs with the considerations of related GDPR articles,
and then manually label 167 privacy policy documents. As each paragraph consists of multiple sentences, we propose 4 different models, which vary on sentence representations and paragraph representations, to learn the paragraph representation from the underlying sentences.
The evaluation results show that the hierarchical model with BERT as sentence representation and BiLSTM as paragraph representation has the best performance, achieving $80.56\%$, and $80.63\%$ in terms of Macro-F1 and Micro-F1, respectively.

To the best of our knowledge, this is the first work to involve the GDPR clauses in privacy policy paragraph classification task. Our key contributions include curating the labeled privacy policy corpus, and benchmarking the paragraph classification task with a novel hierarchical BERT model. Our model has downstream applications, e.g., it could be used for outlining privacy policy with paragraph topical labels, and thus helps users to identify the key information from the long document easier. To spur further research, we will make the labeled corpus and our models publicly available.

\section{Related Work}
\label{sec:relatedwork}
\subsection{Privacy Policy Corpus Creation}
\newcite{PPCorpus16} create a privacy policy corpus (OPP-$115$) of $115$ privacy policies, with crowd-sourcing. They developed a policy scheme with ten data practices, which are
formulated by domain experts. Each data practice further contains multiple  attributes.
An arbitrary length of text span is allowed for the annotations.
The corpus is in relative fine-grained granularity and the evaluation with three classification models shows promising results of automating the labelling process to some extent. Our work targets a different task, which outlines privacy policy with paragraph topical labels.
%
%
%
%
\newcite{VagueCorpus18} create a vague word (word intrinsic property and is irrelevant with the context) corpus for privacy policies, with the targeting task of vague words prediction.
%
%
\newcite{zimmeck2019maps} create an app privacy policy corpus (APP-$350$), targeting the task of App executable and privacy policy compliance checking. They select $350$ policies of the most popular apps from the Google Play Store, and hire legal experts to annotate the data.
\newcite{sathyendra2017identifying} provide a finer-grained corpus based on OPP-$115$~\cite{PPCorpus16} with a semi-automatic labelling process, with the  focus of opt-out choices in privacy policies.
These approaches only cover a small part of privacy policy contents, i.e., APP-$350$ focuses on particular data types that are collected/shared, \newcite{VagueCorpus18} target vague words and \newcite{sathyendra2017identifying} provide labels on opt-out choices. Our corpus focuses on providing topical labels for paragraphs, and covers a wider range of topics compared with these works.

There are also approaches trying to conduct semi-automatic labelling in order to reduce human efforts. \newcite{alignment14} conduct an empirical study on the problem of aligning or grouping segments of privacy polices. Different approaches, i.e., clustering and HMM are studied and compared with manual labelling from Amazon M-Turk. The results show that an automated approach based on word-level similarities could close about half of the gap between automated approaches and median crowd workers.
A recent research~\cite{sarne2019unsupervised} conducts unsupervised learning techniques to study the topics in privacy policies and observes that the topics in the privacy policies, after GDPR is enacted, mismatch the topics in the OPP-$115$ corpus.
\newcite{tesfay2018read} take one step forward to create a corpus including $45$ manually labelled privacy policies. The corpus concentrates on the risk levels of the privacy policies defined by experts.
%

Our work focuses on privacy policy document structure analysis, i.e., we provide topical labels for paragraphs automatically. We reduce this task into a paragraph classification task. We devise a label scheme based on GDPR articles and also consider the topics discovered by the clustering approach~(\newcite{sarne2019unsupervised}) to enrich our label system.

\subsection{Automatic Privacy Policy Analysis}
%
\newcite{liu2018towards} utilize LR, SVM and CNN model to classify the privacy policy segments and sentences, with the corpus created by~\newcite{PPCorpus16}. They further split the 'Other' category of the OPP-$115$ corpus into three categories. i.e., Introductory/Generic, Practice Not Covered, and Privacy Contact Information.
The best results show F1 scores of $0.78$ and $0.66$ for segment and sentence classification, respectively.
%
%
\newcite{kumar2019quantifying}
train a domain specific word embedding with $300K$ privacy policies. They compare the classification results (on three models trained with OPP-$115$) of the  domain-specific word embedding with general word embedding using the GloVe~(\newcite{pennington2014glove}) model. The results show that domain-specific word embedding outperforms general word embedding.
%
\newcite{zimmeck2016automated} propose to combine of machine learning techniques with program static analysis techniques to analyze apps’ potential noncompliance with privacy requirements.
In particular, they adopt the OPP-$115$ corpus to train SVM and logistic regression classifiers, which are used to conduct classification on privacy policies.
%
%
\newcite{chang2019automated} take user profiles into consideration, and automatically list the privacy policy segment description and the corresponding  GDPR descriptions which they predict the users are most interested in. The task is reduced to a segment classification task and the TextCNN classifier is adopted and trained on the OPP-115 corpus to classify each segment of the privacy policy.

Our work targets the document structure analysis of privacy policies, and we reduce the task into a paragraph-level classification task, our label scheme considers the impact of GDPR, which the previous work fails to include.

\section{Task Definition and Classification Scheme Design}
\label{sec:definition}

As reported by~\newcite{cookiestatistics}, the privacy policy contents are largely changed ($72.6\%$ privacy policy updates) after GDPR is enacted. Moreover, there is no existing corpus for analyzing the document structures of privacy policy at the paragraph level. Therefore, we propose to create a labelled corpus to help analyze privacy policy structures.


%
We devise the label scheme based on  GDPR articles, i.e., the articles which define the proper actions that should be reflected in the privacy policy\footnote{Note that GDPR does not explicitly specify which clauses should be reflected in privacy policies, we consult legal domain experts and select the clauses which are objective and suitable to be reflected in privacy policies in our work.}. We also take the OPP-$115$ corpus~\cite{PPCorpus16}, which is a manually labelled corpus for web application privacy policies, as one of the references in our work. Moreover, we adopt the clustering results by~\newcite{sarne2019unsupervised} to help refine our label system.
In particular, the \textit{International Data Transfer}, \textit{Policy Contact Information} are explicitly labelled in our corpus, which are related to \texttt{GDPR Art.13.1(f)} and \texttt{GDPR Art.13.1(a)}, respectively. Some other labels, e.g., \textit{ Cookies and Similar Technologies}, are considered in our label scheme due to their importance and the frequent occurrences in Android privacy policies~\cite{sarne2019unsupervised}.

With the above considerations, we consolidate the topic information~\cite{sarne2019unsupervised} summarized through unsupervised learning techniques, the GDPR regulations as well as and expert knowledge, and propose a label scheme with 11 labels.
%
We introduce the details of our label scheme in the following:
\begin{enumerate}
    \item \textit{Policy Introductory (PI)}: The general descriptions of the privacy policy document, including definitions on the referential pronouns used in the document.
    \vspace{-2mm}
    \item \textit{First Party Collection and Use (FPCU)}: What, when and how the first party (the controller) collects, uses and processes the users' information\footnote{Note that cookies are usually exempted from the category of user information and regarded as the automatically collected information, which is reflected in \textit{Cookies and Similar Technologies}}. [\texttt{GDPR Art.13.1}]
    \vspace{-2mm}
    \item \textit{Cookies and Similar Technologies (CT)}: How to collect and use the cookies and other similar technologies (e.g., beacons), and descriptions about those techniques.
    \vspace{-2mm}
    \item \textit{Third Party Share and Collection (TPSC)}: How the controller shares and discloses the information with third parties,  which include corporate affiliates, service providers or advertising partners.
    \vspace{-2mm}
    \item \textit{User Right and Control (URC)}: The right of user, guaranteed by GDPR and the options which users have in order to control their personal information, such as the settings on users' privacy and safety. For example GPDR requests that the data subject has the right to access, rectify and erase the data. [\texttt{GDPR Art.13.2 (b-f)}]
    \vspace{-2mm}
    \item \textit{Data Security (DS)}: The security facilities/methods that the controller implements to protect users' information. [\texttt{GDPR Art.32.1}]
    \vspace{-2mm}
    \item \textit{Data Retention (DR)}: Descriptions on the retention period about users' information. [\texttt{GDPR Art.13.2 (a)}]
    \vspace{-2mm}
    \item \textit{International Data Transfer (IDT)}: Descriptions of how is the information stored and transferred internationally. [\texttt{GDPR Art.13.1 (f)}]
    \vspace{-2mm}
    \item \textit{Specific Audiences (SA)}: Specific terms for specific audiences, e.g., children, or data subjects from a specific area/country, which usually has privacy protection laws in power. [\texttt{GDPR Art.40.2 (g)}]
    \vspace{-2mm}
    \item \textit{Policy Change (PC)}: Descriptions of changes of the privacy policy and how is the data subject notified if changes happen.
    \vspace{-2mm}
    \item \textit{Policy Contact Information (PCI)}: The contact information of the data controller (i.e., first party). [\texttt{GDPR Art.13.1 (a)}]
\end{enumerate}

\section{Corpus Creation}
\label{sec:annotation}

\subsection{Data Collection}
\begin{table}[t]
\centering
\begin{tabular}{l|c}
No. Documents           & 167     \\
No. Sentences           & 15,614   \\
No. Words               & 447,048 \\ \hline \hline
 Annotated Paragraph          & 5,167    \\
Annotators per Document & 3
\end{tabular}
\caption{The Statistics on the Privacy Policy Corpus}
\label{Tab.ppstatistics}
\end{table}

\begin{table*}[ht]
\centering
\begin{tabular}{l|c|c|c|c|c}
Label                              & Frequency & Coverage  & Avg.S  & Avg.W  & Fleiss' Kappa \\ \hline
Policy Introductory             	&	451	&	0.93	&	2.20	&	51.87	&	0.73	\\
First Party Collection and Use  	&	1647	&	0.99	&	2.58	&	68.95	&	0.71	\\
Cookies and Similar Technologies	&	303	&	0.55	&	3.00	&	66.98	&	0.71	\\
Third Party Share and Collection	&	894	&	0.92	&	2.63	&	71.43	&	0.68	\\
User Right and Control          	&	720	&	0.74	&	2.31	&	54.28	&	0.69	\\
Data Security                   	&	275	&	0.78	&	2.54	&	56.40	&	0.8	\\
Data Retention                  	&	146	&	0.49	&	2.17	&	62.35	&	0.68	\\
International Data Transfer     	&	136	&	0.47	&	2.32	&	63.99	&	0.74	\\
Specific Audiences              	&	214	&	0.70	&	2.71	&	66.01	&	0.76	\\
Policy Change                   	&	176	&	0.76	&	2.81	&	56.22	&	0.74	\\
Policy Contact Information      	&	205	&	0.80	&	1.55	&	28.33	&	0.69	\\

\end{tabular}
\caption{The Per-label Statistics in Our Corpus}
\label{Tab.labelstat}
\end{table*}
Our work focuses on the privacy policy of the Android App. Therefore, we collect privacy policies of Apps from the Google Play\footnote{\url{https://play.google.com/store}}, one of the most popular Android App stores.
We used the Scrapy web framework\footnote{\url{https://scrapy.org/}} and Selenium\footnote{\url{https://docs.seleniumhq.org/}} to automates the data crawling process.
In our work, we aim at collecting a set of high quality privacy policies with a diverse App categories. Therefore, we follow three strategies to collect the seed links: (1) The privacy policies of Apps which are in the top list of Google Play; and (2)
the privacy policies should have a diverse category since different categories may have different requirements on accessing user information.
The privacy policies we collect cover $22$ Google App categories, including  Communication, Game, and Business, etc.
%
After obtaining the seed Android App links, the crawler follows the links to locate the corresponding App and then find the privacy policy link.
%

To ensure the quality of the collected privacy documents, we create the following filtering criteria and the privacy policy documents satisfying all of the criteria are kept: (1) the privacy policy is written in English; and
(2) duplicated privacy polices are removed (some Apps from the same company share one privacy policy); and (3) the content of the privacy policy is of reasonable length. To be specific, we set a minimum size of $2$KB on the privacy policy documents based on observations of average word counts of privacy policies; and (4) the document is describing privacy policy, not some other documents such as Terms Of Services (as some App may put other documents in the link indicating the privacy policy).

We crawl a total of $1,113$ privacy policies from Google Play. After the filtering step with the proposed criteria, $167$ privacy policies remained.
Table~\ref{Tab.ppstatistics} shows the statistics of the privacy policies we labelled.
There are in total $167$ privacy policies labelled, which consists of $5,167$ natural paragraphs and $15,614$ sentences. There is, on average, $32$ paragraphs in each privacy policy document, and each paragraph has around $3$ sentences and $85$ words.
Each privacy policy document is labelled by $3$ annotators.

\subsection{Data Annotation}
Before annotating the data, we conduct pre-processing on the privacy policy document. We remove noises, such as item symbols, header bars of some web pages,  and conduct normalization on links, emails, etc.
We also convert all characters to lower case.

We modify the open source labelling tool named YEDDA~\cite{yang2017yedda} to include our label scheme and enable our labeling task.
In order to properly control the quality of the labelling process as well as the labelled data corpus, we divide our labelling process into two phases.
In the first phase, we ask two master students, who work on related research topics, to label $10$ privacy policies, and then merge the labels to achieve a consensus, during which a discussion of the initial label scheme as well as the labeling process is conducted based on the issues discovered during the labeling process. We then refine our label scheme and process based on the discussion result.
In the second phase, we give a tutorial to all volunteers who are recruited for labelling, which also involve a discussion process to refine the meaning of each label.
%
In total we have $3$ volunteers to label each privacy policy.
To quantitatively measure how the annotators agree on all the labelled sentences, we compute Fleiss' Kappa~\cite{fleiss1971measuring} inter-annotator agreement. As shown in the last column of Table~\ref{Tab.labelstat}, the Fleiss' Kappa values range from $0.68$ to $0.80$, which suggests a substantial agreement among the three raters. To further resolve conflicts, we ask the three volunteers to sit together and discuss the conflicted labels that they provided until a consensus is achieved. The whole annotation process takes a total of $15$ days.

%
%

\subsection{Demographics}
The labelled corpus has a total of $5,167$ paragraphs, and $15,614$ sentences. Table~\ref{Tab.labelstat} shows the statistics of the labelled corpus.
\texttt{Frequency} is the number of data practices (natural paragraphs in our case) appeared in the corpus. \texttt{Coverage} indicates the coverage of the corresponding label, i.e., the percentage of privacy policy documents which contain that label.
We can observe that \textit{First Party Collection and Use (FPCU)} and \textit{Third Party Share and Collection (TPSC)} count for the majority of the paragraphs. This is also consistent with the findings in the OPP-$115$ corpus~\cite{PPCorpus16}. We can also observe that, some of the labels which are designed based on GDPR requirements, such as \textit{User Right and Control (URC)} and  \textit{International Data Transfer (IDT)}, also appear frequently ($74\%$ and $47\%$, respectively) in the privacy policies labelled.
This result indicates that those contents are explicitly documented by many companies (as required by GDPR), and thus the necessity of having those labels.

The \texttt{Avg.S} and \texttt{Avg.W} indicate the average number of sentences and words for each label. We can observe that most topics have an average of $2-3$ sentences describing the contents. Topics on \textit{First Party Collection and Use}, \textit{Third Party Share and Collection} tend to contain more words than other topics. The \textit{Policy Contact Information} label has the smallest number of words.
\section{Privacy Policy Document Structure Analysis}
\label{sec:task}

\begin{figure}[t]
\centering
\includegraphics[width=0.7\textwidth]{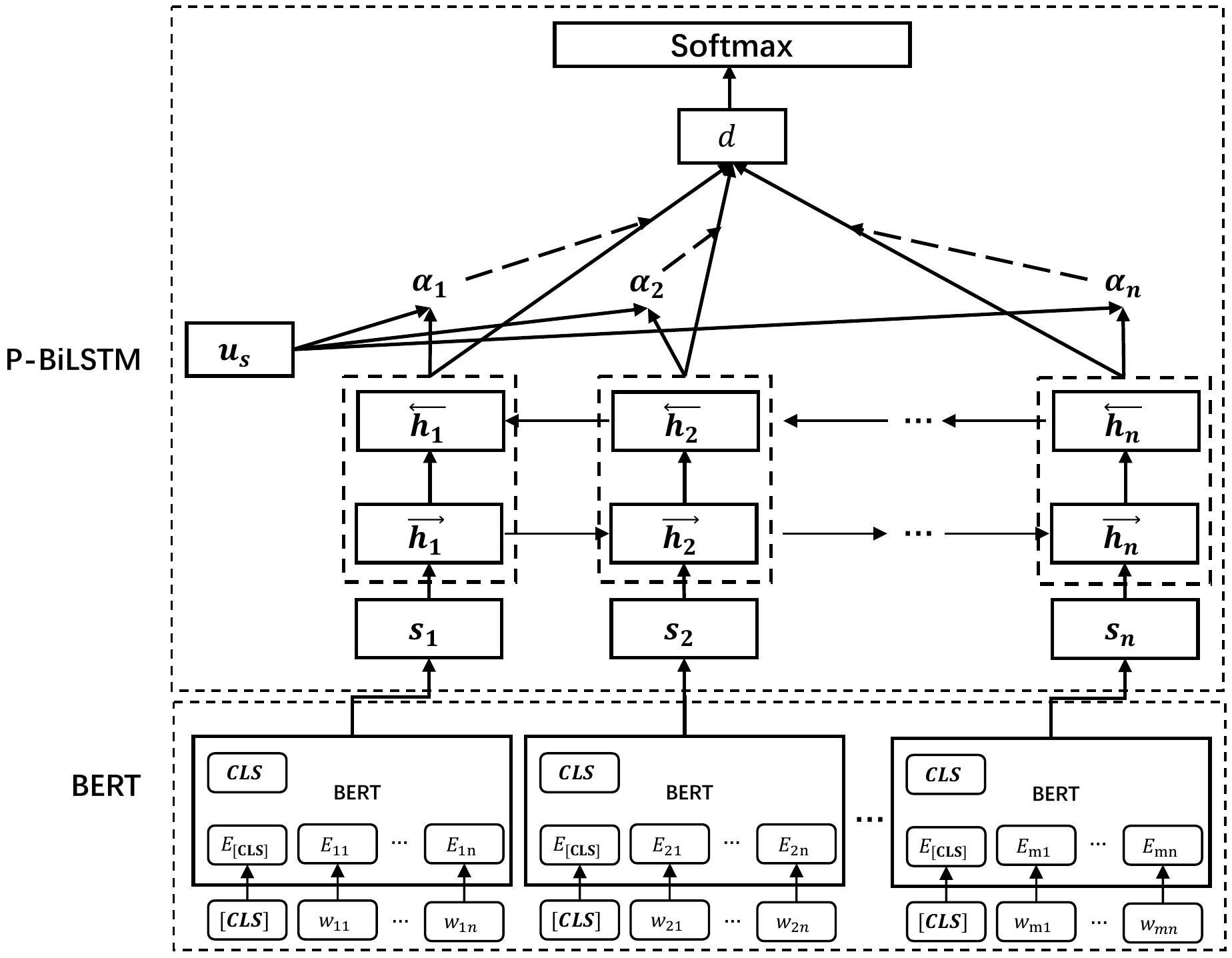}
\caption{The BERT + P-BiLSTM Model Structure}
\label{fig:h-bert}
\vspace{-1mm}
\end{figure}

Our work targets the task of privacy policy document structure analysis. In particular, we frame the task as a paragraph-level multi-class classification problem, aiming at categorizing each paragraph into the 11 pre-defined topical category (detailed in Section~\ref{sec:definition}).
 In this way, we automatically provide a label for each natural paragraph of privacy policies, which could be used for outlining or restructuring long privacy policy documents.

\setlength{\tabcolsep}{3pt}
\begin{table*}[t]
\centering
\begin{tabular}{l|lll||p{1.1cm}p{1.1cm}p{1.1cm}||lll||lll}
\multirow{2}{*}{Label}  & \multicolumn{3}{c||}{S-BiLSTM + P-Attn} & \multicolumn{3}{c||}{S-BiLSTM + P-BiLSTM} & \multicolumn{3}{c||}{BERT + P-Attn} & \multicolumn{3}{c}{BERT + P-BiLSTM} \\
        &   P      &  R      & F        &   P      &  R      &  F    	& P        &  R      & F        & P        & R       & F    	\\ \hline
PI  	&	78.64  &  75.57  &  76.39	&	80.50  &  \textbf{76.42}  &  78.02	&	80.40  &  74.13  &  76.81	&	\textbf{85.89}  &  73.91  &  \textbf{79.02}	\\
FPCU	&	77.38  &  83.81  &  80.38	&	76.57  &  \textbf{85.91}  &  80.85	&	80.29  &  84.06  &  82.01	&	\textbf{82.14}  &  83.88  &  \textbf{82.88}	\\
CT  	&	78.27  &  69.51  &  72.53	&	\textbf{82.99}  &  68.70  &  73.95	&	78.70  &  70.94  &  73.63	&	80.00  &  \textbf{82.99} &  \textbf{76.34}	\\
TPSC	&	72.23  &  78.06  &  74.70	&	77.20  &  78.32  &  77.46	&	\textbf{78.40}  &  79.89  &  \textbf{78.69}	&	75.95  &  \textbf{82.07}  &  78.65	\\
URC 	&	78.53  &  76.73  &  77.22	&	\textbf{78.75}  &  78.08  &  78.02	&	78.43  &  77.56  &  77.70	&	77.54  &  \textbf{80.28}  &  \textbf{78.77}	\\
DS  	&	84.99  &  73.54  &  78.23	&	\textbf{88.62}  &  78.25  &  \textbf{82.24}	&	81.96  &  \textbf{78.40}  &  79.51	&	85.13  &  75.94  &  79.37	\\
DR  	&	\textbf{83.04}  &  64.29  &  71.68	&	78.03  &  70.09  &  73.16	&	75.88  &  \textbf{74.93}  &  74.53	&	81.63  &  74.49  &  \textbf{77.04}	\\
IDT 	&	76.55  &  72.40  &  71.95	&	\textbf{80.77}  &  69.14  &  72.93	&	74.58  &  \textbf{76.49}  &  \textbf{74.58}	&	73.24  &  76.15  &  74.03	\\
SA  	&	89.17  &  80.57  &  84.29	&	89.92  &  80.93  &  84.84	&	85.96  &  84.38  &  84.81	&	\textbf{90.47}  &  \textbf{85.26}  &  \textbf{87.56}	\\
PC  	&	87.50  &  84.14  &  85.50	&	89.10  &  82.24  &  85.05	&	\textbf{90.33}  &  \textbf{87.35}  &  \textbf{88.53}	&	89.47  &  87.13  &  88.04	\\
PCI 	&	84.26  &  67.89  &  74.16	&	\textbf{85.59}  &  63.99  &  72.57	&	76.76  &  75.30  &  75.56	&	81.44  &  \textbf{77.62}  &  \textbf{78.47}	\\ \hline
Micro	&	77.76  &  77.76  &  77.76	&	79.12  &  79.12  &  79.12	&	79.64  &  79.64  &  79.64	&	\textbf{80.56}  &  \textbf{80.56}  &  \textbf{80.56}	\\
Macro	&	80.96  &  75.14  &  77.91	&	\textbf{82.55}  &  75.64  &  78.91	&	80.15  &  78.49  &  79.30	&	82.08  &  \textbf{79.26}  &  \textbf{80.63}	\\

\end{tabular}
\caption{The precision/recall/F score for $4$ classification models}
\label{Tab.classification_result}
\end{table*}

\subsection{Classification Models}
We benchmark the created corpus on the document classification task.
In particular, we adopt $3$ most representative document classification models,
all of which adopt the state-of-the-art deep neural networks to our goal,
where the architecture can be unified as shown in
Figure.
As shown, the overall model architecture can be divided into three parts:
(1) sentence representation, (2) paragraph representation and (3) decoding.
We exploit the most straightforward linear classification as the decoder.
Below we introduce the details of the other two components of our models.

\paragraph{Sentence Representation}
Previously, \newcite{HierarchicalAttention16} have suggested using a bidirectional long short memory neural network (BiLSTM) to encode sentences for paragraph classification.
We exploit this method as a candidate, since their proposed HAN model has been one most representative method for  paragraph classification.
With a sequence of word embeddings as inputs, BiLSTM is applied on the sequences directly, resulting in a new sequence of hidden vectors, which has encoded implicit inter-word relations inside the sentence.
Following \newcite{HierarchicalAttention16}, we then exploit a standard self-attention mechanism to aggregate the sequential hidden vectors into one single feature vector for the final representation. We refer to this method as S-BiLSTM for short.

Recently, the BERT representation~\cite{devlin2018bert}, a kind of deep contextualized word representations, has received great attentions due to its impressive performance on a range of NLP tasks. The BERT representation takes one full sentence as input, and output a sequence of hidden vectors at the word-level, where each vector has encoded the full-sentence information. %
In particular, previous studies usually exploit the outputs of a special token (i.e., [cls]) to represent the input sentence.
We follow the same strategy to obtain the sentence  representation.\footnote{The base, uncased model is exploited in this work, i.e., \url{https://storage.googleapis.com/bert_models/2018_10_18/uncased_L-12_H-768_A-12.zip}}

\paragraph{Paragraph Representation}
When the hidden representation vectors of all sentences in a given paragraph are ready, the next step is to produce one paragraph-level feature vector for decoding.
We use two strategies to achieve our goal.
Following \newcite{HierarchicalAttention16}, our first strategy uses
BiLSTM to enhance the feature composition over sentences, and  self-attention is used for feature aggregation.
The process is essentially the same as that of word-to-sentence composition of the S-BiLSTM method.
For paragraph representation, we denote it by P-BiLSTM for short.
Our second strategy removes the BiLSTM part and uses only the self-attention aggregation layer. This leads to a very concise paragraph representation,  which might be preferable for BERT-based sentence representations.
We use P-Attn to represent this strategy for short.

\paragraph{Final Full Models}
By associating the two components as well as the decoding,
we can obtain four different models, S-BiLSTM + P-BiLSTM (the same as HAN in \newcite{HierarchicalAttention16}), S-BiLSTM + P-Attn, BERT + P-BiLSTM and BERT + P-Attn. The BERT + P-BiLSTM model structure is illustrated in Figure~\ref{fig:h-bert}.

\subsection{Experiment Settings}
We divide the corpus into a training set, a validation set, and a test set according to the ratio of $8:1:1$ (based on privacy policy documents) and  conduct the standard 10-fold cross validation to train the models.
Following the work by~\newcite{kumar2019quantifying}, we pre-trained a  \texttt{FastText}~(\newcite{bojanowski2017enriching}) word embedding model with $20K$ privacy policies crawled from Google Play. The word vector size is set to $100$, the learning rate of is set to $0.25$ and the epoch is set to $10$ following the standard settings.
The \texttt{S-BiLSTM+P-BiLSTM} model uses the pre-trained word embedding as the input vector.
The hidden size of both S-BiLSTM and P-BiLSTM are set to $100$. The batch size is $16$.
For \texttt{BERT}, we used the pre-trained model (BERT-Base, uncased), the batch size is $16$.
For the \texttt{BERT+P-BiLSTM} model, we use BERT to capture sentence representations, and then use the P-BiLSTM structure to capture the paragraph representations.
P-attention is the attention layer which are trained on a randomly initialized context vector.  Each sentence vector (for \texttt{BERT}+\texttt{P-attn}) or hidden vector (for \texttt{BiLSTM}+\texttt{P-attn}) is given a weight $ \alpha $, which is then used to compute the paragraph vector with a weighted summarization on each sentence hidden vector.
As our primary goal is to benchmark the paragraph classification task on our corpus, we adopt state-of-the-art models or structures, e.g., BERT and BiLSTM and self attention mechanisms, which are proved to be effective on document classification tasks. The four models we used are representative on the paragraph classification task.



\begin{figure}[t]
\centering
\includegraphics[width=.6\textwidth]{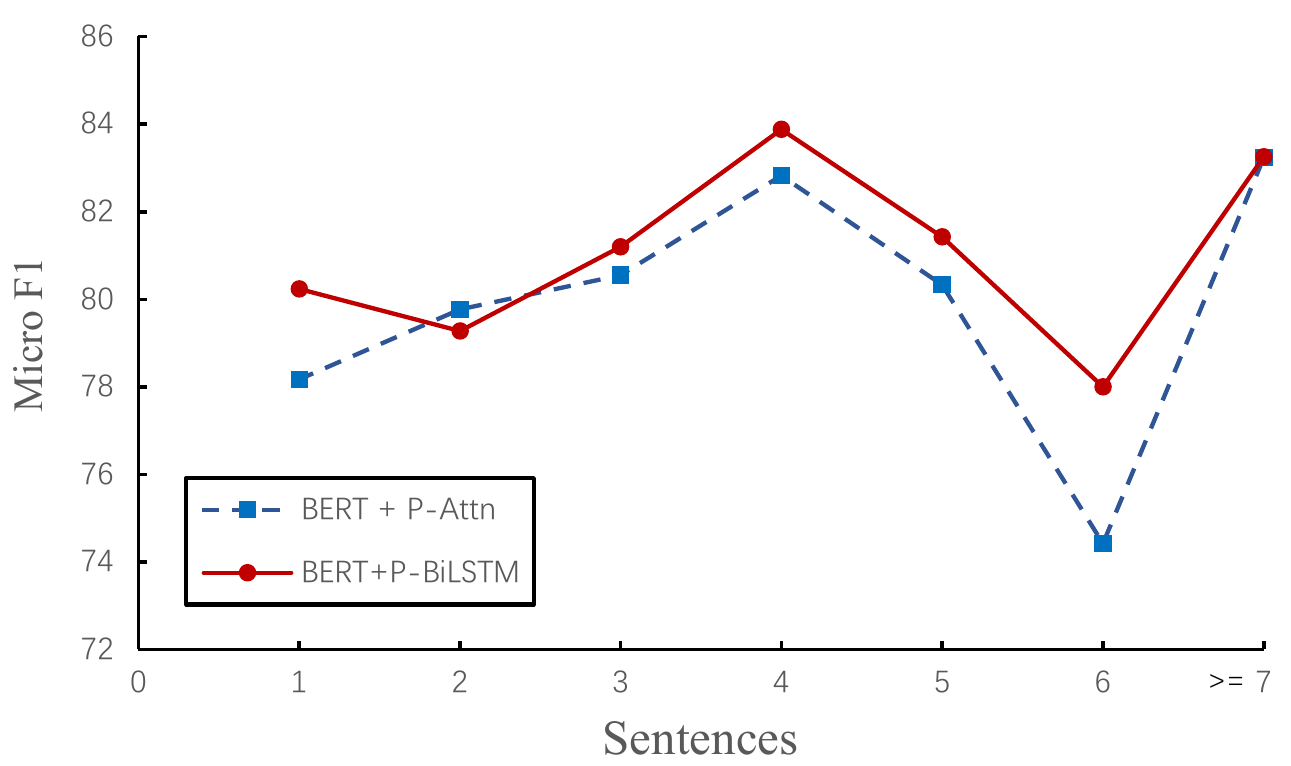}
\caption{F1-measure against Sentence Count for BERT and H-BERT}
\label{Fig.F_vs_S}
\end{figure}



\subsection{Experiment Results}
Table~\ref{Tab.classification_result} shows the classification results, where both Macro and Micro scores are reported. \texttt{P}, \texttt{R}, \texttt{F} represent precision, recall and F1-score, respectively.

We can observe from the results that the BERT + P-BiLSTM model gives the overall best performance on F-scores followed by BERT+P-Attn, which reflects the advantages of a pre-trained model containing semantic information on a small data set.
The models which utilize BiLSTM as sentence encoder (i.e., S-BiLSTM + P-Attn and S-BiLSTM + P-BiLSTM) show worse performance than the models that utilize BERT as sentence encoder  (i.e., BERT + P-Attn and BERT + P-BiLSTM).
The models which utilize the BiLSTM as paragraph encoder (i.e., S-BiLSTM + P-BiLSTM and BERT + P-BiLSTM) show better  performance than those models which utilize only the attention mechanism as paragraph encoder (i.e., S-BiLSTM + P-Attn and BERT + P-Attn).
By a detailed analysis on the classification results, we find that for some cases, the S-BiLSTM + P-BiLSTM model provides high attention to a particular word which is not related to the label, which results in a wrong classification result. For in stance, for sentence \textit{``Please note that despite our efforts, transfer of information on the Internet is never completely secure.''}, the model give the word ``transfer'' a high attention, and label the sentence ``International Data Transfer'', which the true label should be ``Data Secutity'', since the sentence is discussing about data security.
The BERT + P-BiLSTM model shows the best and most stable recall values among all models.

For the \textit{Data Security} category, S-BiLSTM + P-BiLSTM shows the best performance. After analyzing the data, we found that S-BiLSTM + P-BiLSTM performs well in some paragraphs describing encryption methods or protection measures. By further zooming into those data examples, we observe that the average number of sentences in the Data Security category is $2.51$ and there are more than 80\% paragraphs having less than 3 sentences, which means they are short paragraphs. There BERT-based models, i.e.,  BERT + P-Attn and BERT + P-BiLSTM, are not adequately trained (as we will discuss in detail in Section~\ref{sec:analysis}). However, the S-BiLSTM + P-BiLSTM is able to give higher weight to category-specific keywords, e.g.,  \textit{``encrypted''} and \textit{``protect''}, through word-level attention, and finally learns the correct features.


We can also observe that BERT + P-Attn is slightly better than BERT + P-BiLSTM on three labels, i.e., \textit{Third Party Share and Collection} (by 0.04\%), \textit{International Data Transfer} (by 0.5\%), and \textit{Policy Change} (by 0.5\%). By carefully checking the classification results, we find that the total number of test data in those categories are less than 200, which means the testing data in each fold is less than 20. The models tend to show random behaviors in such a small set of test data.





\subsection{Analysis}
\label{sec:analysis}
To evaluate whether the BiLSTM structure improves the classification performance on the same sentence encoder, we plot the F1-score (of BERT + P-Attn and BERT + P-BiLSTM) against number of sentences, and the results are shown in Figure~\ref{Fig.F_vs_S}.
We can observe that the F-score of BERT + P-BiLSTM and BERT + P-Attn shows similar trend with the increase of sentence number and in general BERT + P-BiLSTM has advantage on paragraphs with multiple sentences compared to BERT + P-Attn.
Both models show best F-score with sentence number equals to 4, and worst performance with sentence number equals to 6.
The testing dataset for sentence number equals to 6 is around 10 per fold. Therefore, the results are relatively sensitive to the number of wrongly classified data. The wrongly classified paragraphs may have some entailment of the mis-classified label.
For paragraphs with sentence number to be 4, the categories are among those that have high F-score, e.g., \textit{First Party Collection and Use}, \textit{Specific Audiences} and \textit{Policy Changes}.
BERT + P-Attn performs slightly better (0.50\%) than BERT + P-BiLSTM for paragraphs with $2$ sentences.
By zooming into the details of the wrongly classified data, we find that most of them may contain descriptions that is related to the other labels. For instance, some of the descriptions on \textit{First Party Collection and Use} may contain a few words on \textit{Third Party Share and Collection} and the ``party'' is not clearly stated. BERT + P-BiLSTM may learn rich representation of the paragraph and capture those features. However, the softmax value for the true label is vary close to the (wrongly) predicted label by BERT + P-BiLSTM.

\section{Conclusion and Future Work}
\label{sec.conclusion}

Automatically analyzing privacy policy is an important problem with the increasing concern on human privacy rights. It is thus very critical to create a high quality corpus to assist this task. In this work, we introduce a privacy policy corpus, which consists of $167$ privacy policies. We also benchmark the proposed corpus on the document classification task with $3$ widely-adopted models and a newly proposed model structure. We provide insightful discussions on the evaluation results and provide hints on using the corpus in related research.


Our curated corpus and model have several key applications. Our model can be used to automatically label privacy policies and conduct a large scale of structure analysis. The paragraph label serves as an abstract summary of the corresponding paragraph, which could outline the long privacy policy document, and help users to identify the most important piece of information easier. Since not all privacy policies are well structured, the model could potentially used for facilitating document restructuring.



\bibliographystyle{acl}
\bibliography{main}

\end{document}